\documentclass[aps,prb,twocolumn,groupedaddress,showpacs,amsmath]{revtex4}

\usepackage{graphicx,amssymb}

\begin{document}

%

\title{Magnetic relaxation studies on a single-molecule magnet by time-resolved inelastic neutron scattering}

\author{O. Waldmann}
 \affiliation{Department of Chemistry and Biochemistry, University of Bern, CH-3012 Bern, Switzerland}

\author{G. Carver}
 \affiliation{Department of Chemistry and Biochemistry, University of Bern, CH-3012 Bern, Switzerland}

\author{C. Dobe}
 \affiliation{Department of Chemistry and Biochemistry, University of Bern, CH-3012 Bern, Switzerland}

\author{D. Biner}
 \affiliation{Department of Chemistry and Biochemistry, University of Bern, CH-3012 Bern, Switzerland}

\author{A. Sieber}
 \affiliation{Department of Chemistry and Biochemistry, University of Bern, CH-3012 Bern, Switzerland}

\author{H. U. G\"udel}
 \affiliation{Department of Chemistry and Biochemistry, University of Bern, CH-3012 Bern, Switzerland}

\author{H. Mutka}
 \affiliation{Institut Laue-Langevin, 6 rue Jules Horowitz, BP 156, 38042 Grenoble Cedex 9, France}

\author{J. Ollivier}
 \affiliation{Institut Laue-Langevin, 6 rue Jules Horowitz, BP 156, 38042 Grenoble Cedex 9, France}

\author{N. E. Chakov}
 \affiliation{Department of Chemistry, University of Florida, Gainesville, Florida 32611, USA}

\date{\today}

\begin{abstract}
Time-resolved inelastic neutron scattering measurements on an array of single-crystals of the single-molecule
magnet Mn$_{12}$ac are presented. The data facilitate a spectroscopic investigation of the slow relaxation of
the magnetization in this compound in the time domain.
\end{abstract}


\maketitle

%

A detailed understanding of the magnetic relaxation in small magnetic particles is of fundamental importance
for their application. The discovery of magnetic bistability at the molecular level in polymetallic clusters
such as Mn$_{12}$ac, the so-called single-molecule magnets (SMMs), has opened new possibilities.\cite{Gat03}
The slow relaxation of the magnetization in SMMs below a blocking temperature $T_B$ is associated with a
large ground-state spin $S$ and easy-axis anisotropy $DM^2$ ($S$ and $M$ are the total spin and magnetic
quantum numbers, and $D<0$). The $2S+1$ levels of the spin ground state are thus split, as indicated in
Fig.~2a, producing an energy barrier of height $\Delta \approx -DS^2$ separating the up and down orientations
of the spin (for Mn$_{12}$ac, $S = 10$, $\Delta$ = 67~K, $T_B \approx$ 3.5~K).\cite{Gat03}

The relaxation behavior of SMMs has been studied by various techniques, such as magnetization measurements,
nuclear magnetic resonance, and frequency-domain electron spin resonance.\cite{Gat03,Dre03} In this work are
presented time-resolved inelastic neutron scattering (INS) measurements on an array of single-crystals of
Mn$_{12}$ac. Time-resolved INS has been applied previously to the study of rotational tunneling
phenomena\cite{Muk92}. It is shown here that this approach, due to recent improvements in INS spectrometry,
can be applied to the study of magnetic relaxation.

%

Needle-shaped single crystals of non-deuterated [Mn$_{12}$O$_{12}$(CH$_3$COO)$_{16}$(H$_2$O)$_4$],
Mn$_{12}$ac, with approximate dimensions 5~$\times$~0.5~$\times$~0.5~mm$^3$ and mass ca. 2~mg were obtained
following literature procedures.\cite{Lis80} Mn$_{12}$ac crystallizes in the space group I$\bar{4}$; the
magnetic anisotropy axis $z$ is coincident with the needle axis. This permitted the construction of a sample
consisting of an array of many (ca. 500) single crystals, aligned in grooves (dimensions:
13~$\times$~0.6~$\times$~0.6~mm$^3$, 16 grooves per platelet) milled into 1~mm thick 15~$\times$~15~mm$^2$
aluminium platelets (Fig.~1a). A total of 15 platelets were stacked in an aluminium container (Fig.~1b), and
sealed under He with an indium seal and aluminium screws.

Inelastic neutron scattering (INS) measurements were performed on the time-of-flight spectrometer IN5 at the
Institut Laue-Langevin in Grenoble, France. The sample was inserted into an orange $^4$He cryostat equipped
with a 2.5~T vertical-field magnet. The sample was aligned with the $z$ axes of the crystals parallel to the
magnetic field. The field was changed at the maximum ramp rate provided by the power supply (4~mT/s for
fields above ca. 0.2~T, but significantly slower for lower fields). An initial neutron wavelength of
5.9~{\AA} was used, affording a resolution of 60~$\mu$eV at the elastic line. In this energy range, the
background of aluminum is negligible. The data were corrected for detector efficiency using a vanadium
standard. The spectra shown below correspond to the sum over all the scattering angles.

\begin{figure}
\includegraphics{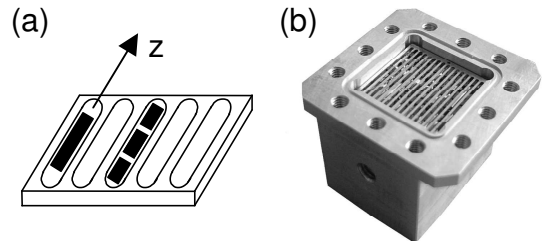}
\caption{\label{fig1} (a) Schematic of an aluminium platelet. The black bars depict Mn$_{12}$ac crystals
positioned in the grooves, such that the anisotropy axes $z$ (= needle axes) are oriented along the grooves.
(b) Photograph of the aluminium box containing the platelets filled with Mn$_{12}$ac crystals.}
\end{figure}

%

In the following the dependence of the Mn$_{12}$ac INS spectrum on a magnetic field $B$ applied parallel to
the $z$ axis shall first be discussed. Measurements were made by first cooling the sample in zero field from
5~K, i.e., above the blocking temperature, to 1.5~K. Accordingly, the lowest levels on both sides of the
double-well potential, i.e., the $M = \pm 10$ levels, were equally populated, see Fig.~2a (the population
densities of the $M=\pm10$ levels are denoted as $n_\pm$). The field was then increased stepwise. The
resulting spectra are shown in Fig.~2d. The zero-field spectrum shows the known peak at $E_0$=1.24~meV,
corresponding to the two transitions $M = -10 \rightarrow -9$ ($I_-$) and $M = +10 \rightarrow +9$
($I_+$),\cite{Mir99,Bir04} as indicated in Fig.~2a. Increasing the field leads to a tilting of the
double-well potential due to the Zeeman energy, see Fig.~2b. The energies of the two transitions $I_-$ and
$I_+$ are now different, they appear at $E_- = E_0 + g \mu_B B$ and $E_+ = E_0 - g \mu_B B$, respectively ($g
= 2.0$, $\mu_B$ is the Bohr magneton). The experimental data for $B \leq$ 0.5~T indeed exhibit two peaks at
exactly $E_-$ and $E_+$, see Fig.~2d. Note that for the fields used here the level $M = +10$ lies
considerably above the ground state, e.g., $20 g \mu_B B$ = 0.93~meV at 0.4~T, and would thus normally be
depopulated. The observation of both transitions $I_-$ and $I_+$ thus directly reflects the slow relaxation
of the magnetization in Mn$_{12}$ac at 1.5~K, which permits the metastable state with the $M = +10$ level
populated. For $B >$ 0.5~T, however, the system completely relaxed back to the ground state due to the
crossing of the $M=+10$ and $M=-9$ levels (Fig.~2c). Accordingly, only the $I_-$ peak was observed in this
field range. Interestingly, the linewidths are resolution limited, indicating an alignment of the crystals
within 5$^\circ$. This is possibly because of a self alignment of the crystals due to the torque exerted by
an applied magnetic field.

\begin{figure}
\includegraphics{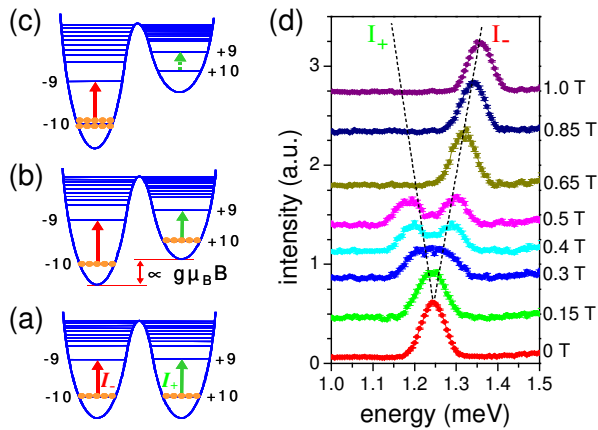}
\caption{\label{fig2} (a-c) Schematics showing the transitions observed for (a) $B = 0$, (b) $0 < B \leq$
0.5~T, and (c) $B >$ 0.5~T. Transitions $I_-$ and $I_+$ are indicated by arrows. Circles symbolize the
populations of the $M = -10$ and $M = +10$ levels. (d) 1.5~K neutron-energy-loss spectra for the magnetic
fields indicated. Curves are offset for clarity.}
\end{figure}

%

The above experiments clearly show i) that for $B \leq$ 0.5~T it is possible to create a metastable situation
of the populations and ii) that for $B > 0$ the populations of the $M = +10$ and $M = -10$ levels can be
measured individually via the intensities of the transitions $I_-$ and $I_+$ appearing at different transfer
energies. This permitted a time-resolved study of the magnetic relaxation in Mn$_{12}$ac as follows: First,
only the $M = +10$ level was populated by applying a negative field of -0.85~T. Then, the field was swept
quickly to a positive value $B > 0 \leq 0.5~T$, so that all molecules were in the now higher lying $M = +10$
level. This triggered the relaxation of the system towards the $M = -10$ level, which was followed by
repeated INS measurements (a run time of $\Delta t$ = 5~min already gave reasonable spectra; typically,
however, $\Delta t$ = 30~min was chosen). In Mn$_{12}$ac the relaxation time is some years at 1.5~K, and a
few seconds at the blocking temperature. Thus, through varying the temperature, the relaxation time could be
tuned such that time slices of $\Delta t$ = 5-30~min allowed reasonable time resolution. Specifically,
time-resolved INS spectra were recorded for $T$ = 2.45, 2.65, and 2.70~K with $B$ = 0.4~T (data not shown).
At $B$ = 0.4~T, the $I_-$ and $I_+$ transitions appear at transfer energies of $E_-$=1.29~meV and $E_+$ =
1.19~meV, respectively, see Fig.~2d.

A more refined measurement protocol, denoted stop-and-go (s\&g), allows for an improvement over the above
constant-$T$ protocol. Here, the temperature is set to a value $T^*$ much lower than the temperature $T$ at
which the relaxation is to be studied, such as to essentially freeze the system. After initialization steps
as above (but at $T^*$), the relaxation is followed by repeatedly setting the temperature to the target
temperature $T$, waiting for a time $\delta t$, cooling back to $T^*$, and measuring for a time $\Delta t$.
Since the relaxation time at $T^*$ is much longer than at $T$, this amounts to taking snapshots of the
population distribution at intervals of $\delta t$. The possibility of freezing a state during recovery to
equilibrium thus permits both short time intervals and reasonable acquisition times. The time-resolved INS
spectrum was recorded with the s\&g protocol at $T$ = 2.65~K with $\delta t$ = 2~min, see Fig.~3a ($B$ =
0.4~T, $T^*$ = 2.3~K, $\Delta t$ = 30~min; an animated movie is provided in the supporting material).

\begin{figure}
\includegraphics{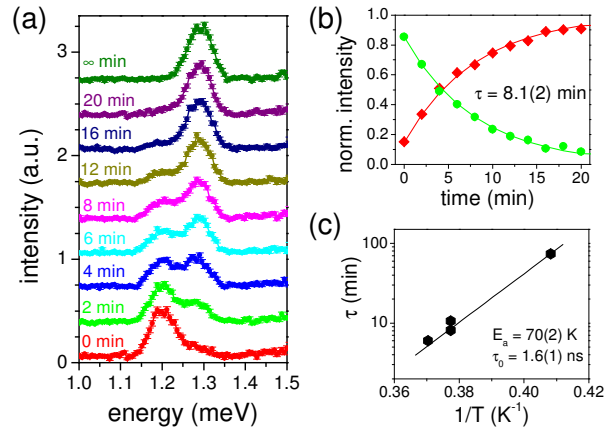}
\caption{\label{fig3} (a) Neutron energy-loss spectra at 2.65~K as measured with the s\&g protocol (see
text). Curves are offset for clarity. (b) Time dependence of the normalized intensities of the transitions
$I_-$ (squares) and $I_+$ (circles). Solid curves correspond to a fit to an exponential decay with a
relaxation time $\tau$ as indicated. (c) Arrhenius plot for the relaxation times. Circles are experimental
points, the line is a fit to the Arrhenius law $\tau = \tau_0 e^{E_a/(k_BT)}$ with parameters as indicated.}
\end{figure}

The experimental data clearly allow one to follow the relaxation of the populations of the $M=\pm10$ levels.
The time dependence of the peak intensities were analyzed quantitatively by fitting the data at each time
step to two Gaussians with a constant offset. In the fits, the linewidths of the Gaussians were fixed to the
spectrometer resolution. The values obtained for the peak intensities were then normalized such as to add up
to 1 on average. The result is shown in Fig.~3b. The time dependence of the peak intensities, or populations
$n_-$ and $n_+$, respectively, are well described by an exponential decay, $n_-(t) = \exp(-t/\tau)$ and
$n_+(t) = 1 -n_-(t)$, as demonstrated by the solid lines in Fig.~3b. For the current experiments, the
population was not inverted completely after the initialization step, i.e., about 15\% of the molecules
remained in the ground state (this is evident from the 0~min curve in Fig.~3a, which shows small intensity
also at 1.29~meV). This is attributed to the slow ramp rates provided by the magnet power supply near zero
field.

The relaxation times obtained from the exponential decay fits were 8.1(2)~min for the s\&g experiment at
2.65~K, and 74(2), 10.6(7), and 6.0(5)~min for the constant-$T$ experiments at 2.45, 2.65, and 2.70~K,
respectively. The relaxation times obtained by the s\&g and constant-$T$ protocols at 2.65~K are in good
agreement. Plotting the relaxation time as a function of the inverse temperature (Arrhenius plot, Fig.~3c),
reveals a thermally activated behavior consistent with thermally-assisted tunneling of Mn$_{12}$ac in this
temperature range.\cite{Gat03} The activation energy $E_a$ = 70(2)~K, thus determined, is slightly larger
than the zero-field value of 62~K.\cite{Gat03} Zero-field relaxation measurements on SMMs generally yield
$E_a < \Delta$, which has been explained by efficient tunneling paths near the top of the
barrier.\cite{Gat03} In the present case, however, the Zeeman energy removes these paths increasing $E_a$.

\begin{figure}
\includegraphics{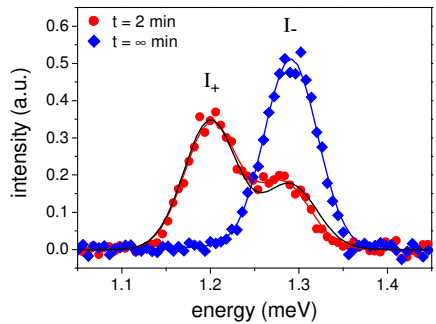}
\caption{\label{fig4} INS spectra after 2~min and long waiting times recorded at 2.65~K using the s\&g
protocol. Solid lines are best-fit curves, for details see the text.}
\end{figure}

The data, allowing also for a time-resolved investigation of the peak positions, revealed an interesting
effect. After initialization, both peaks were shifted slightly towards the center $E_0$ as compared to their
equilibrium positions $E_\pm = E_0 \pm g \mu_B B$. The shifts were small, about 8~$\mu$eV, but statistically
significant. This is illustrated by the s\&g-INS spectra of Fig.~3a at 2~min and after long waiting times,
presented in more detail in Fig.~4. The two solid lines which fit the data well correspond to best-fit
curves, where both intensities and positions were fitted. The third curve, which approximates the 2~min data
with less accuracy, was obtained from a fit but with the transfer energies fixed to the equilibrium values.
The shift of the $I_-$ peak at ca. 1.29~meV towards lower energies is clearly visible. The observations may
be described in terms of a time-dependent effective magnetic field $B(t) = B_{eq} - \Delta B(t)$, which
differs from the field at equilibrium $B_{eq}$. The experiments yield $\Delta B(0) \approx$ 70~mT. As a
function of time, the peak positions approached their equilibrium values, but the limited statistics of the
data did not allow a precise determination of the relaxation law. The effect described has been observed in
all measurements with similar values for $\Delta B(0)$. It might seem natural to associate $\Delta B(t)$ with
the magnetization $M = B - H$ (in units of T). However, for needles the depolarization factor $N = 1$ and the
essentially cubic arrangement of the Mn$_{12}$ac molecules in the crystals means that the local magnetic
field is equal to the external magnetic field. Further studies are necessary to elucidate the details of this
phenomenon.

%

In conclusion, we have demonstrated that with high-performance time-of-flight spectrometers, such as IN5 at
the ILL, the magnetic relaxation processes in magnetic compounds can be studied with time-resolved INS
measurements. We believe that the spectroscopic resolution provided by this methodology will allow
unprecedented insights into the physics of magnetic relaxation.

%

\begin{acknowledgments}
Financial support by EC-RTN-QUEMOLNA, contract n$^\circ$ MRTN-CT-2003-504880, and the Swiss National Science
Foundation is acknowledged.
\end{acknowledgments}

%

%
\end{document}